\documentstyle[prl,aps,twocolumn,epsf]{revtex}
\def \be {\begin{equation}}
\def \ee {\end{equation}}

\begin{document}

\title{Random Matrix Approach to Glassy Physics --- Low Temperatures and Beyond}
\author{Reimer K\"uhn and Uta Horstmann}

\address{Institut f\"ur Theoretische Physik, Universit\"at Heidelberg, Philosophenweg 19, 69120 Heidelberg, Germany\\{\rm (Submitted 9 Jan 97)}}

\address{~
\parbox{15cm}{\rm
\medskip
A random matrix approach to glassy physics is introduced. It leads to a 
class of models which exhibit both, glassy low--temperature phases, and 
double-- and single--well configurations in their potential
energy. The distribution of parameters characterizing the local 
potential energy configurations can be {\em computed\/}, and  
differ from those assumed in the standard tunneling model and its variants. 
Still, low--temperature anomalies characteristic of amorphous systems
are reproduced, and we are able to distinguish properties which can be
expected to be universal from those which cannot.
\\{~}
}}

\maketitle

\narrowtext

Glassy  materials exhibit, at low temperatures, a number of properties which 
are considered {\em anomalous\/} in comparison to those of their crystalline 
counterparts\cite{huphi}. Examples are the roughly linear temperature 
dependence of the specific heat $C(T)$ \cite{zepo} and the approximately 
quadratic temperature variation of the thermal conductivity $\kappa(T)$ 
\cite{fran} at $T < 1$\,K, which are in contrast to the $T^3$ behaviour of 
these quantities in crystals. Moreover, between approximately 1 and 20\,K, 
one observes a crossover to a $T^3$ behaviour of the specific heat and a 
plateau in the thermal conductivity. The anomalies below 1\,K appear to 
be {\em universal\/} in the sense that they are shared by a large variety of 
amorphous systems, whereas between approximately 1 and 20\,K a stronger 
dependence on properties of the specific material appears. At still somewhat 
higher temperatures, a certain degree of universality is again observed,
at least in thermal conductivity data\cite{fran}.

It is the existence of a broad range of localized low energy excitations
which is generally held responsible for the aforesaid
anomalies. The dominant mechanism giving rise to excitations below 1\,K is
believed to be tunneling in double--well potentials (DWP). This is the 
content of the phenomenological standard tunneling model (STM) \cite{an+,phil}. A different set of excitations in amorphous systems can exist as localized 
vibrations in soft anharmonic single--well configurations of the potential 
energy. They have been postulated within the, likewise phenomenological 
soft--potential model (SPM) \cite{kar+,bu+} to describe the physics also 
above 1\,K. Either model requires specific assumptions concerning the distributions of its parameters \cite{an+,phil,kar+} to describe glassy physics 
at low $T$.

The {\em existence\/} of localized soft vibrations \cite{schol},
and DWPs \cite{heusi} responsible for two--level tunneling systems (TLS) 
has been convincingly established in molecular dynamics studies of glassy 
systems. In \cite{heusi}, the {\em statistics\/} of the parameters 
characterizing the local potential energy configurations was also 
investigated (within the confines of a generalized SPM). Still, as of 
now it is perhaps safe to characterize both, STM and SPM as providing 
{\em phenomenological\/} descriptions, based on assumptions which, while 
plausible in many respects, are still lacking support based on more 
microscopic approaches, such as that of \cite{gra+} for KBr:KCN mixed crystals. 
At the same time they are widely used to explain and fit experimental 
results. In particular, it has been argued \cite{yule} that the STM could not 
really account for the considerable degree of universality observed in glassy 
physics, and numerical investigations on specific substances 
\cite{schol,heusi} can at best only partially elucidate this point. It was 
suggested \cite{yule} that universality in glassy systems would emerge only 
due to sufficiently long--range {\em interactions between\/} TLSs, an idea 
which has been taken up in \cite{bur+}.

Recently, one of us has proposed a random--matrix model of a glass, which 
exhibits both, an amorphous low--temperature phase, and a collection of 
double-- (and single--)well configurations in its zero--temperature potential 
energy surface \cite{ku96}. The main motivation for that study was to formulate 
a bona--fide model of an amorphous system for which the statistics of local 
potential energy configurations becomes part of the world of the {\em 
computable\/}. The result, taken as input for computing the influence of 
tunnelling--excitations on the low--temperature properties, was used to 
demonstrate that the model shares the low--temperature anomalies characteristic 
of glassy systems. Moreover, it was shown that one could also exhibit {\em 
relations\/} between low--$T$ and high--$T$ phenomena, e.g. between the 
low--temperature specific heat and the value of the glass--transition 
temperature itself, and relate this, in turn, with features of the system 
at the microscopic level of description.

The purpose of the present paper is, {\em inter alia\/}, to investigate
the model beyond the two--level approximation in the quantum--mechanical 
treatment of its local excitations. Thereby we find that it reproduces ---
{\em without\/} further assumptions --- also the characteristic bump as it 
is observed in $C(T)/T^3$ plots of the specific heat. We can  trace this 
back to {\em microscopic\/} features of our model, and we are able to 
address the universality question posed in \cite{yule}: Our answer,
in short, is that the ensemble of local potential energies is
itself a largely collective affair, and some degree of universality
may therefore be expected to arise {\em without\/} interactions at the level
of quantized local excitations. We shall elaborate on these points below,
and also try to situate our model in glassy physics, beyond the specific 
assumptions embodied in it.

Let us restate the main ingredients of our model. It is based on the following 
ansatz for the potential energy of $N$ degrees of freedom (``particles") 
forming a glass--like system,
\be
U_{\rm pot}(v) = -\frac{1}{2} \sum_{i\ne j} J_{ij} v_i v_j + \frac{1}{\gamma}
\sum_i G(v_i)\ ,
\ee
in which $v_i$  may be interpreted as the deviation of the 
\mbox{$i$-th} particle from some preassigned position. The amorphous 
aspect is modelled by taking the first, harmonic contribution to $U_{\rm pot}$ 
to be {\em random\/}, so that the reference positions would generally turn out 
to be unstable at the harmonic level of description. A set of anharmonic 
on--site potentials $G(v_i)$ is therefore added to stabilize the system as a 
whole. We choose the harmonic part such that the model can be analysed exactly 
within mean--field theory, and the replica method is used to deal with the 
disorder. Namely, we take the $J_{ij}$ to be Gaussians with mean $J_0/N$ and 
variance $J^2/N$, specializing to $J=1$ to fix the energy--scale. For the 
on--site potential we choose
\be
G(v) = \frac{1}{2} v^2 + \frac{a}{4\,{\rm !}} v^4\ .
\ee
That is, $G$ also creates a harmonic restoring force, and by varying $\gamma$ 
we can tune the number of modes in the system which are unstable in the harmonic approximation. Other forms of $G(v)$ may be contemplated; our ability to solve 
the model does not depend on the particular shape of $G$. The only requirement 
is that it increases faster than $v^2$ for large $|v|$ for the system to be 
stable, and our choice is the simplest respecting a $Z_2$ symmetry.

The choice of random couplings in (1) puts our model outside the class of 
glass--models in the narrow sense. In view of recent ideas concerning the 
fundamental similarity between quenched disorder and self--induced disorder as 
it is observed in glassy systems proper \cite{mez++}, it may still be 
argued that our choice should capture {\em essential aspects\/} of glassy 
physics. Models of this type with different couplings and different choices 
for $G(v)$ have been studied in the context of analogue neuron systems 
\cite{ho,kub}.

In a mean--field setting, the system is described by an ensemble of effective 
single--site problems, characterized by potentials which, in replica theory, 
attain random parameters. Solving the model means to compute the distribution 
of parameters characterizing the local potential energy configurations 
self--consistently, and this is precisely what we were afer: it can directly 
be translated into the distribution of the parameters characterizing DWPs. 
Taking this distribution as input of a tunneling model, we were able to {\em 
compute\/}, for instance, the contribution of the tunneling states to the 
specific heat at low $T$ \cite{ku96}. In what follows, we shall go beyond the
two--level tunneling approximation, and begin to explore the consequences.

To analyze the potential energy surface, one computes the (configurational)
free energy of the system
\be
f_N(\beta) = -(\beta N)^{-1} \ln \int \prod_i d v_i \exp[-\beta U_{\rm 
pot}(v)]
\ee
and takes its $T=0$ limit to eliminate entropic contributions, {\em and\/} 
to select one of the system's (many) ground state configurations, using 
replica theory to average over the disorder so as to get {\em typical\/} 
results. Standard arguments \cite{SK} give $f(\beta) = \lim_{n\to 0} 
f_n(\beta)$ for the quenched free energy, with
\begin{eqnarray}
n f_n(\beta) & = &\frac{1}{2} J_0 \sum_a p_a^2 + \frac{1}{4}\beta \sum_{a,b} 
q_{ab}^2 \nonumber \\
& - & \beta^{-1} \ln \int \prod_a d v^a \exp\big[-\beta U_{\rm eff} (\{v^a\})\big]\ .
\end{eqnarray}
Here
\be
U_{\rm eff}  = - J_0 \sum_a p_a v^a - \frac{\beta}{2} \sum_{a,b} 
q_{ab}v^a v^b  + \frac {1}{\gamma}\sum_a G(v^a)
\ee
is a replicated single--site potential and the order parameters $p_a =N^{-1} 
\sum_i {\langle v_i^a\rangle}$ and $q_{ab} =N^{-1} \sum_i {\langle v_i^a v_i^b\rangle}$ are determined from the fixed point equations
\begin{eqnarray}
p_a & = &\langle v^a \rangle \quad ,\ a=1,\dots ,n \\
q_{ab} & = &\langle v^a v^b \rangle \quad ,\ a,b = 1,\dots ,n\ ,
\end{eqnarray}
in which $\langle\dots\rangle$ denotes a Gibbs average corresponding to the 
replica potential (5), and the limit $n\to 0$ is eventually to be taken.

We have evaluated (4)--(7) in the replica symmetric (RS) \cite{ku96} and the
1st step replica--symmetry breaking (1RSB) approximations. In RS one assumes
$p_a = p$ for the `polarization', and $q_{aa}=\hat q$ and  $q_{ab}=q$ for $a\ne 
b$ for the diagonal and off--diagonal entries of the Edwards-Anderson matrix. These are determined from
\begin{eqnarray}
p  = \langle\, \langle v\rangle\, \rangle_z\ , \  \
C  = \Big\langle\, \frac{1}{\sqrt q}\frac{d\,\langle v\rangle}{d z}\, 
\Big\rangle_z\ , \ \
q   =   \langle\, \langle v\rangle^2\, \rangle_z\ .
\end{eqnarray}
Here  $C=\beta (\hat q - q)$, and $\langle\dots\rangle_z$ denotes an average 
over a standard Gaussian $z$ while $\langle\dots\rangle$ is now a Gibbs average 
corresponding to the RS single--site potential
\begin{eqnarray}
U_{\rm eff}(v)  = - h_{\rm RS}\, v -\frac{1}{2} C v^2 + \frac{1}{\gamma} G(v)\ ,
\end{eqnarray}
with $h_{\rm RS} = J_0 p + \sqrt{q}\, z$. The RS approximation thus describes 
a Gaussian ensemble of single-site potentials $U_{\rm eff}(v)$, with parameters 
$p$, $q$ and $C$ which are determined self--consistently through (8).

The system described by (8)--(9) exhibits a transition from an ergodic phase 
with $p=q=0$ to a glassy phase with $q\ne 0$ at some temperature $T_g$ 
depending on $J_0$ and $\gamma$; see Fig.~1. If $J_0$ is sufficiently large, a 
transition to a polarized phase with $p\ne 0$ may also occur. The assumption 
of RS is not always correct. Replica symmetry breaking (RSB) occurs at low 
temperatures and large $\gamma$. The location of the instability against RSB is 
given by the AT criterion \cite{AT} $1=\beta^2 \, \left \langle\,(\,\langle 
v^2\rangle - \langle v\rangle^2\,)^2\, \right \rangle_z$. In the 1RSB 
approximation, which is expected to constitute a major step towards the full
solution, $U_{\rm eff}(v)$ is of the same form (9), albeit with $h_{\rm RS}$ replaced by $h_{\rm 1RSB} =  J_0 p + \sqrt{q_0}\, z_0 + \sqrt{q_1-q_0}\, z_1$, and $C$ by $C=\beta (\hat q - q_1)$, where we use standard notation \cite{SK} 
for the entries of the $q$--matrix. Along with a so--called partitioning 
parameter $m$, they are determined from a more complicated set of fixed point 
equations \cite{SK,hoku}. Now $z_0$ is a Gaussian, whereas the distribution of 
$z_1$ is more complicated.

\begin{figure}[t]
\centering
  \epsfysize=5cm
  \epsffile{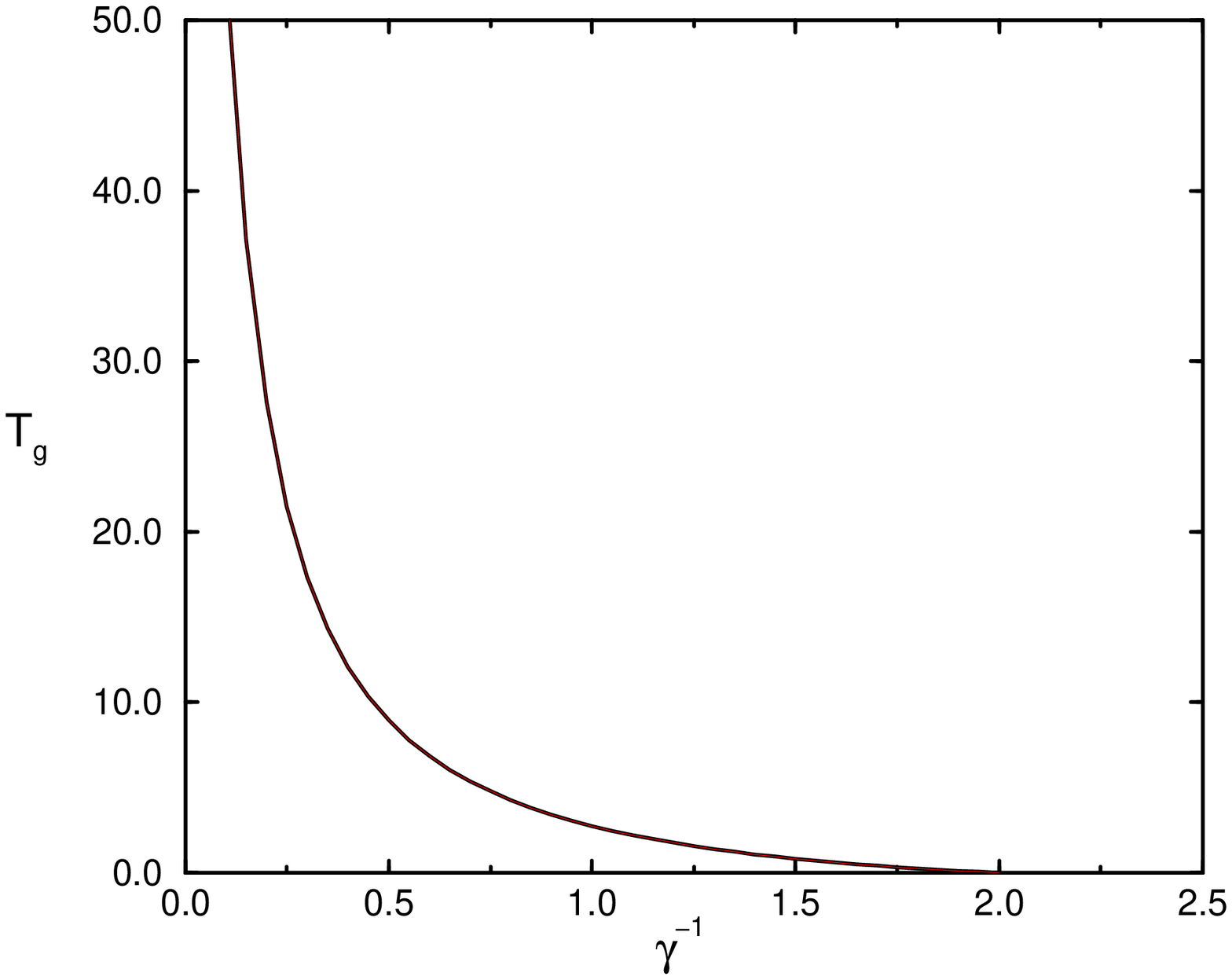}
\centering
 \epsfysize=5cm
\epsffile{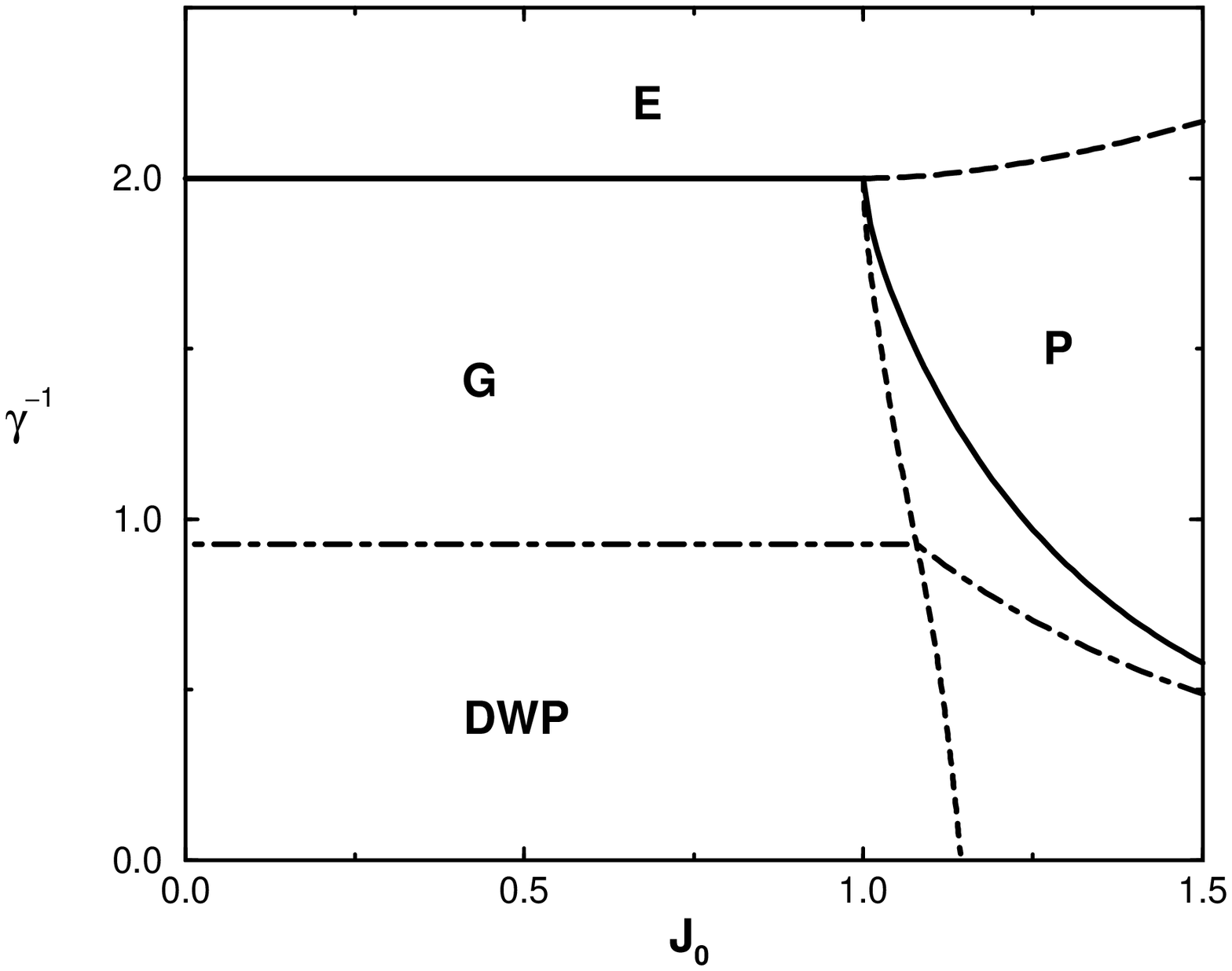} 
\caption[]{{\bf (a):} Glass temperature $T_g$ as a function of $\gamma$ for 
$J_0=0$. {\bf (b):} $T=0$ phase diagram in RS. $E$ denotes the $T=0$--limit of 
the ergodic phase, $G$ the glassy phase, and $P$ a phase with macroscopic 
polarization. The full line is the AT line. Below the dot--dashed line is 
the region with DWPs. The short--dashed line separates the glassy phase $G$ 
from the phase $P$ with macroscopic polarization.}
\end{figure}

Interestingly, the system can also exhibit a collection of DWPs in its $T=0$ 
potential energy surface. That is, for suitable external parameters, some of 
the $U_{\rm eff}(v)$ have DWP form. From (2),(9) it is clear that the 
condition for this to occur is $\gamma C > 1$; this region is marked DWP in 
Fig.~1b. DWPs exist for not too large $|h_{\rm RS/1RSB}|$, and the distribution 
of their characteristic parameters derives from the Gaussian nature of $z$ (the 
joint $(z_0,z_1)$--distribution in the 1RSB case) and can be {\em computed}. 
These are barrier height $V$, asymmetry $\Delta$, distance $d$ between the 
wells, and the tunnelling matrix element $\Delta_0 = \hbar \omega_0 \exp 
(-\lambda)$, with $\lambda=d (2 m_0 V / \hbar^2)^{1/2}$, $\omega_0$ 
a characteristic frequency (of the order of the frequency of harmonic
oscillations in the two wells forming the DWP) and $m_0$ the effective mass of 
the tunneling particle. For larger $| h_{\rm RS/1RSB}|$, the $U_{\rm eff}(v)$ 
only exhibit single--well forms. 

In contrast to the main assumption of the STM that $\Delta$ and $\lambda$ are 
uncorrelated with $P(\Delta,\lambda) \simeq P_0$, we find $\Delta$
and $\lambda$ to be strongly correlated; both are functions of one random 
variable $h_{\rm RS/1RSB}$ in RS/1RSB respectively, a feature which persists at 
all finite levels of RSB. By the same token, SPM expansion--coefficients
of $U_{\rm eff}(v)$ are correlated. The correlations between $\lambda$ and 
$\Delta$ and likewise those between the SPM coefficients can be weakened (but 
{\em not\/} eliminated) by introducing {\em local\/} randomness, i.e., by 
making either $\gamma$ or other parameters of the on--site potentials 
$i$--dependent in a random fashion. It has been demonstrated in \cite{kub} 
that our model remains solvable with these modifications. The RS distributions 
$P(\lambda)$ and $P(\Delta)$ have been presented in \cite{ku96}. Both have 
singularities at their upper boundary, the former an integrable divergence, 
the latter a cusp singularity. 

\begin{figure}[h]
\centering
  \epsfysize=5cm
  \epsffile{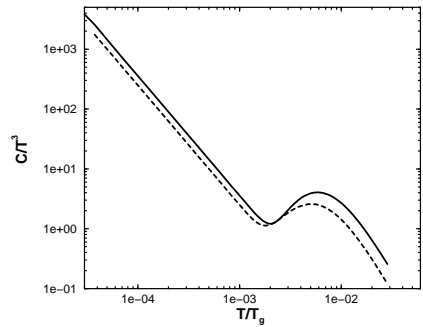}\hfill
\caption[] {Low $T$ specific heat in the $G$--phase for $\gamma = 10$, and 
$J_0=0$ in the RS (full line) and 1RSB (broken line) approximation. Note that 
the location of the bump at $T_b \simeq 6\times 10^{-3} T_g$ is reasonable for
many glasses. E.g. for a-SiO${}_2$ this would correspond to $T_b\simeq 8.6$\,K.}
\end{figure}

Even though $P(\Delta,\lambda)$ is different from what is assumed in the STM, 
we find that the contribution of the tunneling excitations to the specific heat  exhibits an extended range of temperatures where it scales linearly with $T$. 
By solving the Schr\"odinger equation of a particle moving in the potential 
$U_{\rm eff}(v)$ and by computing the specific heat {\it beyond\/} the 
two--level approximation, we obtain a bump in a $C(T)/T^3$--plot (Fig.~2), as 
it is typically also observed in experiments. Its {\em microscopic\/} origin is 
related to ``harmonic" excitations in DWP and SWP structures whose frequency 
depends only weakly on $h_{\rm RS/1RSB}$, thus $\Delta$, and which gives rise 
to a peak in the density of states (DOS). For DWPs with larger asymmetries, the 
role of tunneling is, indeed, mainly to prevent level crosssings of these 
``harmonic" excitations as $\Delta$ is varied.

Within our model there is a clear distinction between the {\em local\/} $G(v_i)$
contributions to $U_{\rm pot}$ and the {\em global\/} ones mediated by the
$J_{ij}$. It is the influence of the latter which makes the ensemble of local 
potential energy configurations a {\em collective\/} affair and which can thus
be regarded as a source of universality of glassy low--temperature physics.
The effect of the $J_{ij}$ on the $U_{\rm eff}(v_i)$ is to produce a linear
term $-h_{\rm RS/1RSB}\, v$, which creates a broad range of asymmetries and
thereby a nearly constant contribution to the DOS at low energies, and  a 
renormalization of the harmonic contribution through $-\frac{1}{2} C v^2$, 
which is responsible for the very occurrence of DWPs. Conversely, local 
properties embodied in $G(v)$ are mainly responsible for the ``harmonic"
excitations creating peaks in the DOS {\em on top\/} of the unstructured part.
These cannot be expected to be universal, which fits rather nicely with the
experimental observation that both the bump in the $C(T)/T^3$ plot and the
corresponding plateau in $\kappa(T)$ (presumably created by resonant scattering 
of phonons from these harmonic excitations) do show a stronger material 
dependence. It is not unreasonable to assume that such a distinction between 
local and global contributions to $U_{\rm pot}$ can be made for real materials 
as well (perhaps with randomness in the local contribution), the long--range 
global one being, e.g., of elastic origin. Interestingly, on introducing local 
randomness such as non--degenerate distributions for the $\gamma$, we have 
obtained slightly super-linear low--T specific heats, with exponents ranging 
between 1 and approximatly 1.5  \cite{hoku}, which fits nicely with 
experimental observations. It is currently under systematic study. As mentioned 
above, local randomness will also weaken the perfect correlation between 
$\lambda$ and $\Delta$ which is also desirable in view of the experimental 
situation \cite{huphi}. 

From the nearly constant DOS at low energies we can determine the constant
$P_0$\cite{huphi}. For parameters as in Fig.~2, we find $P_0\simeq 6\cdot 
10^{-3} (k_{\rm B}\, K)^{-1}$/atom in RS (aproximately half this value in 
1RSB), which is roughly three orders of magnitude too large for most glasses. 
While this may be regarded as a deficiency, we believe that it does not 
invalidate our approach (no tuning of parameters whatsoever has yet been 
attempted) but rather that it may be used to guide further detailed 
modelling: again, local randomness turns out to be a possible cure.

We have proposed and solved a simple model which exhibits both, an amorphous
low--temperature phase, and a collection of DWPs and SWPs in its  potential 
energy landscape, and which reproduces characteristic glassy low temperature 
anomalies, though, as a model based on localized degrees of freedom it cannot 
be expected to present a faithful picture in all details of the ergodic (liquid) phase. Also, we do not presently regard our model as providing a realistic 
description of specific substances but rather as representative for a new and 
promising way of looking at glassy physics --- at low temperatures and beyond. 
For instance, models of the type proposed will also  exhibit interesting {\em 
dynamical\/} properties at or near their glass temperature $T_g$ and possibly 
throughout their glassy phase, which might be investigated, e.g., along the 
lines of \cite{sozi}. Finally, we believe that our approach does hold the 
potential for eventually including realism; the search for good models has only 
just begun.
 
We would like to thank C. Enss, H. Horner, S. Hunklinger, J. van Mourik,
P. Neu, O. Terzidis, J. Urmann, and A. W\"urger for  numerous illuminating discussions.

\end{document}